\documentclass[10pt,oneside]{article}

\usepackage[numbers]{natbib}
\usepackage{subfigure}
\usepackage{dsfont}
\usepackage{pdfpages}
\usepackage{algorithm, algpseudocode}
\usepackage{algcompatible}
\usepackage{booktabs,makecell,tabularx}
\usepackage{amsmath,amssymb,amsfonts}
\usepackage{graphicx}
\usepackage{textcomp}
\usepackage{xcolor}
\usepackage{bbm}
\usepackage{tabularx}
\newcommand{\algmargin}{\the\ALG@thistlm}
\makeatother
\algnewcommand{\parState}[1]{\State%
    \parbox[t]{\dimexpr\linewidth-\algmargin}{\strut\hangindent=\algorithmicindent \hangafter=1 #1\strut}}
\usepackage{tikz}
\usetikzlibrary{trees}
\newcommand{\etal}{\textit{et al}.\textrm{ }}
\tikzstyle{bag} = [align=center] 

\algdef{SE}[SUBALG]{Indent}{EndIndent}{}{\algorithmicend\ }%
\algtext*{Indent}
\algtext*{EndIndent}
\def\BibTeX{{\rm B\kern-.05em{\sc i\kern-.025em b}\kern-.08em
    T\kern-.1667em\lower.7ex\hbox{E}\kern-.125emX}}

\begin{document}



\title{MRCN: Enhanced Coherence Mechanism for Near Memory Processing Architectures}  
\author{Amit Kumar Kabat, Shubhang Pandey and TG Venkatesh} 
\maketitle

\begin{abstract}
In Near Memory Processing (NMP), processing elements(PEs) are placed near the 3D memory, reducing unnecessary data transfers between the CPU and the memory. However, as the CPUs and the PEs of the NMP use a shared memory space, maintaining coherency between them is a challenge. Most current literature relies on maintaining coherence for fine-grained or coarse-grained instruction granularities for the offloaded code blocks. We understand that for most NMP-offloaded instructions, the coherence conflict is low, and waiting for the coherence transaction hinders the performance. We construct an analytical model for an existing coherence strategy called CONDA, which is within 4\% accuracy. This model indicates the key parameters responsible - the granularity of offloaded code, probability of conflicts, transaction times, and commit time. This paper identifies the prospective optimizations using the analytical model for CONDA. It proposes a new coherence scheme called MRCN: Monitored Rollback Coherence for NMP. MRCN addresses the coherence issue while eliminating unnecessary re-executions with limited hardware overhead. The MRCN is evaluated on synthetic as well as Rodinia benchmarks. The analytical results are within 4\% accuracy of the simulation results. The MRCN shows improvement of upto 25\% over  CONDA strategy for the same benchmark under different execution conditions.

\textbf{Keywords: }Near Memory Processing, 3D stacked memory,  High Bandwidth Memory, Coherence Mechanism, Performance analysis.
\end{abstract}



 

\section{Introduction}
The  performance of the processor has continued to improve over the past few decades because of the concepts of pipelining, on-caches, chip multiprocessor architectures, and the application of many more strategies [12]. However, the  performance of DRAM has failed to grow at the same pace as the processor. The  performance of the processor is greatly affected by the time spent accessing the data. Further off-chip memory accesses can consume almost half of the total energy and time needed for computation [2]. The growing difference in the performance of the processor  and the performance of the main memory is known as the memory wall problem [12]. Modern-day applications ranging from machine learning, Image Processing, to Graph Applications are becoming extremely data-intensive, thereby aggravating the memory wall problem. 
With the success in 3D fabrications, researchers have found computing closer to memory as a possible solution to the memory wall problem [19]. Currently, products like Hybrid Memory Cube (HMC) [22] and High Bandwidth Memory (HBM) [18] enable computing near the memory. The 3D memory structures generally involve DRAM dies stacked one above another and are connected using Through Silicon Via (TSV).  The highlighting feature of the 3D memories is the TSVs which can provide bandwidth as large as 420GB/s [22].

There are two main approaches to computing closer to the memory. They are (i) Processing-in-Memory (PIM) and (ii) Near Memory Processing. Processing-in-Memory (PIM) [19]  is one approach where the characteristics of the memory bits are used, and changes in the circuit level of memory are done to perform certain operations and are the fastest way possible to execute certain operations. However, the PIM approaches are specific to the needs of the application. So, to make the system more general and support a broader range of applications, Near Memory Processing (NMP) has been proposed [23]. In NMPs, the bottom-most layer is available for the Processing Element (PE) that enables the computation. Alternatively, PEs are placed by the side of the 3D memory. The PE used in NMP can also be specialized, targeting a single application if necessary. The programs running in PE benefit from the high internal bandwidth of 3D memory using TSV to access the data. This eliminates the excess off-chip transfer required for the CPU.  

\begin{figure}[!htp]
    \centering
    \includegraphics[width=0.75\textwidth]{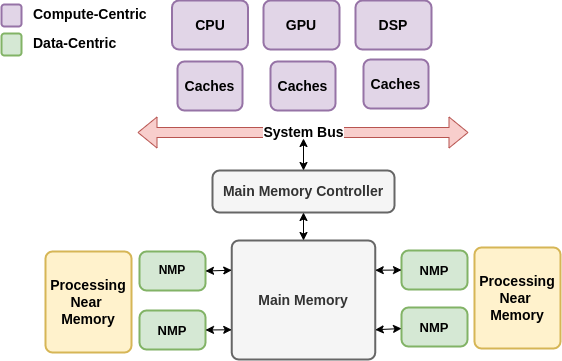}
    \caption{High-level diagram of Near Memory Processing architectures}
    \label{fig:NMP_arch}
\end{figure}

Performing computation closer to the memory has many advantages, such as, faster memory accesses and highly parallel execution. However, a major challenge in NMPs is the maintenance of the coherence between the CPU and the PE [25]. Although the PE of NMP and the CPU operate independently, they share the same memory space. Thus it is highly possible that both processing units access the same data leading to data inconsistency. Fig. \ref{fig:NMP_arch} presents a high level diagram of the NMP architectures, depicting both the NMP PE and CPU operating over a shared memory space. Given the separate locations of the PE and the CPU operating at different speeds, maintaining coherence is a considerable challenge. Depending on the coherence approach followed by the NMP, the performance of the whole system can vary drastically  [25]. As a result new unconventional methods for maintaining coherence between CPU and NMP are needed to address this problem. Motivated by this issue, we set the goal of the paper as to propose an enhanced coherence mechanism between CPU and NMP, so as to improve the system performance. 

The organization of the rest of the paper is as follows. In section \ref{background}, we briefly discuss about the features of the 3D stacked memories and the prevailing coherence mechanisms used in the multi-core architectures. Section \ref{relatedworks} focuses upon the coherence strategies used in the NMP architecture development. The analytical modelling of some of the existing strategies and possible optimizations are presented in section \ref{analyticalmodel}. Section \ref{methodology} presents the approach used in performing this study. Validation of the study and performance comparison between the proposed coherence strategy and the existing strategy is discussed in section \ref{results}. Finally, the conclusion is presented in section \ref{conclusion}.

\section{Background} \label{background}

This section briefly discusses the Near Memory Processing (NMP) architecture, features, and constraints. We provide background on the existing coherence methods observed in multi-core architectures and coherence approaches applied for NMPs. 


\subsection{Background on 3D Stacked Memory}
\begin{figure}[!htp]
    \centering
    \includegraphics[width=\textwidth]{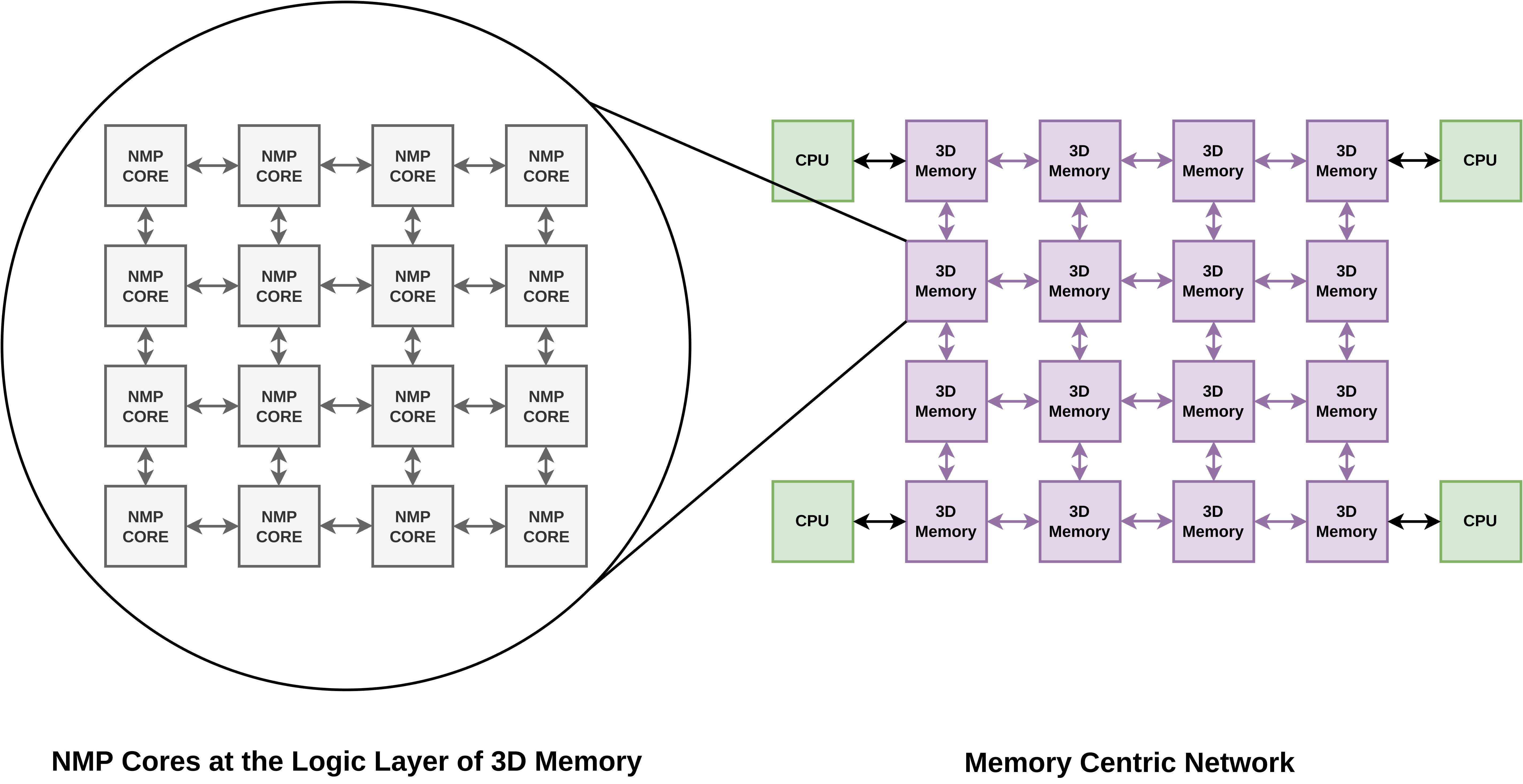}
    \caption{Baseline diagram of the NMP CPU architecture}
    \label{nmp_baseline_diagram}
\end{figure}

Increasing bandwidth and memory density has been a constant challenge for designers. From DDR to DDR4, the bandwidth of a single memory package increased from 2.66 GB/s to 21.3 GB/s while energy consumption per bit was reduced by nearly six-fold [26]. Then HMC by Micron increased it to 320 GB/s  [22] due to the use of TSV. HBM [14] from Samsung used TSV, Silicon interposers, and micro bumps to create a 2.5D structure having a bandwidth of nearly 256 GB/s. HBM and HMC offered the additional option to employ NMP by placing processing units at the bottom layer of the 3D structure. The structure eliminates the whole unnecessary data movement. However, due to area constraints and thermal restraints of a 3D structure, a powerful host CPU cannot be used as an NMP PE. So depending on the PEs used in the bottom layer, categorically selected programs or portions of programs are offloaded to the NMP section [15]. 
The baseline diagram for our NMP architecture is shown in fig. \ref{nmp_baseline_diagram}

\subsection{Background on Coherence Mechanism}
In the CPU-NMP system, the  PEs of the NMP benefits more than the conventional CPU cores due to their proximity to the memory. The complete CPU-NMP arrangement must be treated like any other multi-core system. So maintaining communication between the  PEs of NMP and the CPU cores along with their respective application execution is an additional challenge. The  approaches adopted to solving the coherence problem can be divided based on the granularity of the task. Granularity is  defined as the size of the task assigned to a core in a multi-core system [12]. As the granularity changes, the communication overhead between cores also changes. The coherence problem can also be viewed in two ways in terms of granularity [12], i.e., (1) Fine-grained and (2) Course grained.

Fine-grained coherence is the traditional one that uses cache block, the finest granularity possible for coherence. On each cache miss, the core  looks for the data in the shared cache of other cores, ask for validation and then perform the execution. So the coherence is maintained on the instruction level. Suppose the data is not present even in other cores, then it goes for the DRAM resulting in frequent communications. In [4], Cantin \etal has mentioned that on average, 67\% (which can go up to 94\%) of these communications are unnecessary. The fine-grained coherence can be efficient in a system without NMP as cores, as the multi-cores are placed on a single chip, and communication is fast. However, fine-grained coherence is a massive disadvantage in an NMP system where the host core and PE are placed at different locations, and the entire broadcast would be done through off-chip channels.

In the second approach, the coarse-grained coherence reduces frequent communication between cores by monitoring the coherence status of a more significant portion of memory instead of only a single cache line. However, this also adds complexity in terms of hardware, depending on the size of the portion to be validated. Increasing the channel size between cores to fit more memory information is challenging. 

Nevertheless, in NMP, supported by 3D-stacked memories, the communication between CPU cores and PE will be done through comparatively large off-chip channels. For example, compared to a single-channel GDDR5 of 32 bits, HBM provides eight 128-bit channels [16]. So coarse grained coherence can be a viable option for NMP.

\subsection{Tasks \& Selection of Offloading Tasks} \label{nmp_task_offload}

We have used an OpenMP-based parallel programming model for the shared memory system [7]. The tasking model in OpenMP allows us the controllability in maximizing the benefits from parallelism. The granularity of offloading the task depends on the benchmark which is being evaluated. Traditional synchronization primitives are to be used by the programmer for the task execution. For example, in the ferret benchmark, the main application is to perform the image similarity search. To perform it, we have different pipeline stages. The five pipelined stages are defined as the individual tasks. There is no data dependency for different images, and the pipelining can be performed easily. However, for individual operations, the respective data dependencies must be maintained [5]. 

We follow the same offloading strategy as proposed in [1]. Ahn \etal [1] presented that the tasks within an application which are capable of exploiting locality more should be executed on the CPU side rather than the PIM side. The conventional out-of-order CPUs are more powerful than the in-order NMP cores. The cache size is smaller on the NMP side, which would generate more conflicts by default. Despite the higher performance of the CPU core, having tasks that depend less on cache support would deteriorate the system performance because of the excessive stalls due to the cache misses. Such tasks which do not heavily rely on locality can be executed parallelly. It is suggested that these tasks are computed near the memory such that they take advantage of the high memory bandwidths and limit the time spent in memory accesses. Therefore, we offload the tasks by observing the L1 cache miss status as we profile the benchmarks.



\section{Related Works} \label{relatedworks}

This section will survey the literature on current coherence mechanisms and strategies proposed for PIM/NMP. Cache coherence in PIM architectures is challenging as it is tough to maintain coherence between the host CPU and the PE in a memory unit. Conventional protocols like MESI (Modified-Exclusive-Shared-Invalid)
[9], MOESI (Modified-Owned-Exclusive-Shared-Invalid) [8], MESIF (Modified-Exclusive-Shared-Invalid-Forward) [11] are suitable for multi-core processors sharing cache on a single chip. 


Gagandeep Singh \etal [25] have highlighted the importance of having a robust Coherence Mechanism for maximizing on the benefits proposed by the NMP architecture. A more positive way is to divide the program to execute the partitions in CPU and PIM/NMP. The works by Pattnaik \etal [21], Gao \etal [10], and Hsieh \etal [13] assume that the partitions do not access the same data region or may access the same data regions at different times. However, restricting CPU or PIM to a particular data region limits the system performance. Partitioning the programs into PIM and CPU without sharing the data between them is a challenging task, and the researchers in this domain are giving much effort. 

Some of the existing course-grained solutions which do not have all these constraints are CONDA, an extension of LazyPIM [2,3]. They allow the PE in memory to perform in parallel with the host CPU without sending large off-chip messages for coherence. The NMP/PIM cores execute on a data set, assuming no coherence conflict. After the execution is completed, they send a compressed signature containing all the addresses used during the process. If any conflict exists, NMP cores roll back and re-execute that set.  

While operating on the enormous bulk of data with numerous PEs on the NMP side, most current literature relies on fine-grained or coarse-grained strategies. We realize that for most NMP offloaded instructions, the coherence conflict is low, and waiting for the transaction hinders the performance and reduces the acquired advantages. Under this optimism, we consider that the NMP side has full access to the memory it is operating. However, it does not commit the data until the coherence transaction is completed between the PEs and the CPU. We analytically model an existing strategy- CONDA [2,3], which helps us identify potential optimizations for further improvement. Realizing the potential optimizations, we propose an enhanced coherence model MRCN (Monitored Rollback Coherence for NMP). The unique features of our work are mentioned as follows.

\begin{enumerate}
\item Our work is the first to provide an analytical model for the coherence strategies in the NMP architectures. The model allows us to understand better the existing coherence communication between the CPU cores and the NMP PE.

\item With nominal hardware overhead, our work proposes an enhanced coherence mechanism-Monitored Rollback Coherence for NMP. The study shows that the MRCN achieves significant improvement in execution cycles with reduced effort in re-execution in case of a coherence conflict.  

\item Not only have we studied the performance of the system against synthetic benchmarks. But also, the performance of the existing coherence strategies and the proposed coherence strategy is evaluated against the real work benchmarks from the Rodinia Benchmarks Suite. 

\end{enumerate}

The major contributions of our work can be stated as follows.

\begin{enumerate}
\item We analytically model the existing coherence model used in the NMP architectures. The analytical model results in execution cycles lie within $4\%$ accuracy, even when subjected to worst-case conditions.

\item We identify the critical parameters in coherence transactions that contribute to the added time in execution. These critical parameters can be listed as -the granularity of offloaded code, probability of conflicts, transaction times, and commit time. These parameters help us propose a new coherence mechanism - the MRCN strategy.

\item We formulate an analytical model for the MRCN strategy as well, the accuracy of which also lies within $4\%$ accuracy. The MRCN strategy outperforms the CONDA strategy by $5\%-25\%$ when tested under the same environment. 

\end{enumerate}

\section{Analytical Modelling and MRCN} \label{analyticalmodel}

In this section, we will first model the existing strategies [2,3], and formulate an equation representing the average execution time. Later, we identify the optimization knobs from the equation developed for CONDA and propose improvements. The last part of this section will present an equation for average execution time, when improvements are done.

\subsection{Coherence Mechanism}

\begin{figure}
    \centering
    \includegraphics[width=\textwidth]{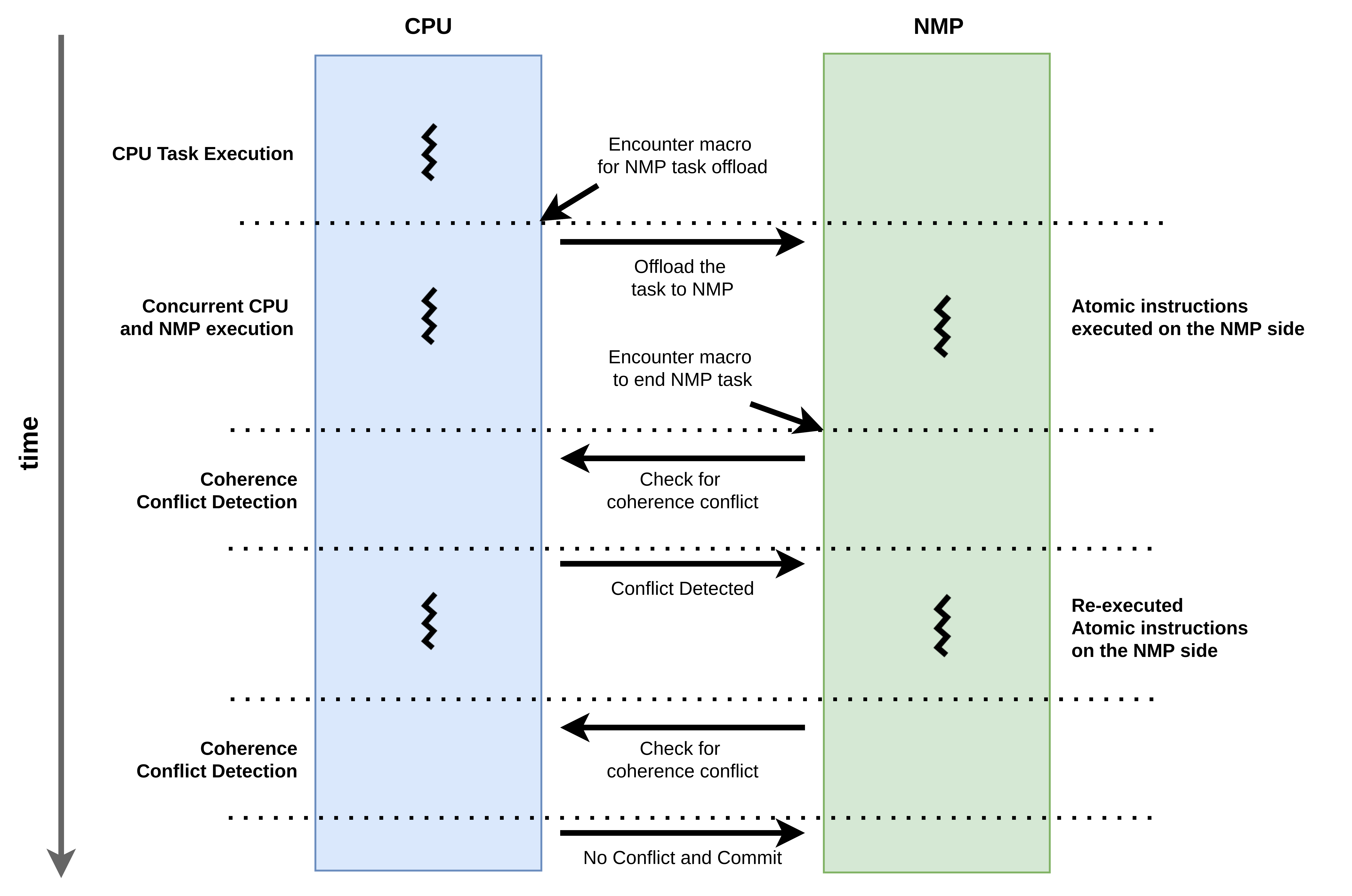}
    \caption{Coherence Mechanism in [2,3]}
    \label{fig:my_label}
\end{figure}

The functionality of the NMP system can be divided into three sections (1) NMP side execution, (2) CPU side execution, and (3) the coherence communication between NMP and the host CPU. The NMP side performs execution speculatively under the optimism for reduced coherence conflict, just like LazyPIM [2] and CONDA [3]. The NMP assumes that it has all the coherence permissions needed for the operation. Only after the execution is done a compressed message containing the reads and writes done during the execution is sent to the host side for coherence validation. On the CPU side, the validation is done by comparing the addresses sent from NMP to CPU writes done during that period. 

Following the work [2,3], the following cases do not require the NMP to re-execute its offloaded section: Cases \textbf{CPU Read and NMP Write} and \textbf{CPU Write and NMP Read}. 
NMP side waits till the conflict report is back. Once the message is received, if there is any conflict, i.e., there was a match found during validation, then the NMP re-executes the task again in the next slot ( Boroumand \etal , [2,3] ).

\subsection{Formulating the CONDA strategy} \label{formulating LazyPIM}

The NMP and CPU side operate on shared memory segment. We consider $\mathcal{K}$ to be the number of memory address that can be held in the shared space, where any memory address can be accessed with uniform probability $\frac{1}{\mathcal{K}}$. Let $f_{cpu}$ and $f_{nmp}$ be the fraction of the  shared memory segment that the CPU and the NMP side accesses  for the execution of the allocated tasks.

For the model, let us assume that $I$ denote as the total number instructions to be executed on the NMP side. These instructions can be grouped into $N$ blocks, where each block can be identified as $B_i$, for $i=1,2, \dots N.$ Let $\theta_i^{nmp}$ denote the number of instructions within each offloaded block $B_i$. Therefore, $I=\sum_i \theta_i^{nmp}.$ 

While execution, the NMP side assumes that it has all the access to the memory and only when the execution of offloaded instructions is complete the coherence check would be made.  Let $\mathcal{N}_i$ denote the memory addresses accessed by the NMP during the execution of the offloaded block $B_i$, and $|\mathcal{N}_i|$ is equal to $f_{nmp} \times \theta_i^{nmp}$. Let $\theta_i^{cpu}$ denote the number of instructions executed on the CPU side before the coherence transaction is received from the NMP. A set $\mathcal{C}_i$ denotes the memory addresses accessed by the CPU in the shared memory space before the next coherence transaction is received, and $|\mathcal{C}_i|$ is equal to $f_{cpu} \times \theta_i^{cpu}$. 

For $r=1,\dots \mathcal{K},$ let $X_r$ be a random variable, such that $X_r = 1$ when both the CPU and the NMP side access the same memory address and $X_r = 0$ otherwise. Probability of atleast one conflict is equal to 1 minus the probability of no conflicts. The probability for memory address not being used from the NMP side can be written as $(1-\frac{1}{\mathcal{K}})^{|\mathcal{N}_i|}$ and similary for the CPU side can be given as $(1-\frac{1}{\mathcal{K}})^{|\mathcal{C}_i|}$. Hence,

\begin{equation}
    P(X_r=1)=(1-(1-\frac{1}{\mathcal{K}})^{|\mathcal{N}_i|})(1-(1-\frac{1}{\mathcal{K}})^{|\mathcal{C}_i|})
\end{equation}
    
\begin{equation}
    \therefore, \;\; P(X_r=0)=1-(1-(1-\frac{1}{\mathcal{K}})^{|\mathcal{N}_i|})(1-(1-\frac{1}{\mathcal{K}})^{|\mathcal{C}_i|})
\end{equation}

The probability of conflict for each block of instructions offloaded can be defined as $P_i^c.$

\begin{equation}
    P^i=P(\sum_r X_r \geq 1)=1-[P(X_r=0)]^\mathcal{K}
\end{equation}
Finally, 
\begin{align}
\label{LazyPIM_conflict_prob}
    P^i=1-[1-\{1-(1-\frac{1}{\mathcal{K}})^{|\mathcal{N}_i|}\}\{1-(1-\frac{1}{\mathcal{K}})^{|\mathcal{C}_i|}\}]^\mathcal{K}
\end{align}

Let $T_{inst}$ denote the average execution time for each instruction. Let, $T_{tran}$ be the transaction time between the NMP and the CPU side which is basically the time taken to send the coherence message (signature) from NMP to CPU and receive the response for the same. On an average 40-50 cycles are taken to send the signature containing address list and receive back the coherence report [3]. Additional 8 cycles are needed to commit the changes from the NMP side which is represented by $T_{commit}$. It is appropriate to mention that the CPU always commits the changes to the memory. However, the NMP will only commit to these changes once the coherence transaction is complete.

Now, if there are no coherence conflicts between the CPU and the NMP, then the total execution time for the offloaded instructions in $B_i$ can be expressed as,

\begin{align}
\label{no conflict LazyPIM}
    T_{\substack{\text{no coherence}\\\text{conflict in} B_i}}= \theta_i^{nmp} T_{inst} + T_{tran} +  T_{commit}
\end{align}

Each instruction on average takes $T_{inst}$ time to execute, and there are $\theta_i^{nmp}$ instructions in the offloaded instruction block $B_i.$ Thus, average execution time to execute all the instructions would be equal to $\theta_i^{nmp} T_{inst}$. Once, execution is completed the coherence transaction would be initiated and as mentioned earlier $T_{tran}$ time would be consumed. Currently, we have assumed that no coherence conflict occurs so the memory commit from the NMP side happens and it consumes $T_{commit}$ amount of time.

In the next case, we will assume that a coherence conflict occurs in the block $B_i$, and identify the average execution time of $B_i$ in which the conflict has occurred. The initial flow will be the same until the first coherence transaction is made. The time spent until then is given as $\theta_i^{nmp} T_{inst} + T_{tran}$. Let $\alpha$ denote $\theta_i^{nmp} T_{inst} + T_{tran}$. The coherence transaction will now indicate that a conflict has occurred and the all $\theta_i^{nmp}$ instructions in block $B_i$ will be re-executed and a coherence transaction will be made. Finally, the memory commit will be performed. Therefore the total execution time for the offloaded instructions in $B_i$, when a single coherence conflit is observed can be expressed as,
\begin{align}
\label{no conflict LazyPIM single execution}
    T_{\substack{\text{with  
    coherence}\\\text{conflict in} B_i}}= 2\alpha +  T_{commit}
\end{align}

Since, we have defined the probability of conflict as the probability that both the CPU and the NMP will access the same memory segment during the same period. Then, the average memory execution time for any block $B_i$, having $\theta_i^{nmp}$ instructions can be defined as,

\begin{align} \label{LazyPIM_exe_time}
  \mathbb{E}[T_i]  & =  P^i \{2\alpha +  T_{commit}\} \\ \nonumber
   &\;\;\;\;\;\;\;\;\;\;\;\;+(1-P^i)\{\alpha +  T_{commit}\}\\ \nonumber
   & =\alpha(1+P^i) +  T_{commit}
  \end{align}

The formulation resembles the approach followed in the work [2,3]. The average total execution time for the strategies in [2,3], can be expressed as $\mathbb{E}[!htp]=\sum_i^N   \mathbb{E}[T_i]$. 

\subsection{MRCN: Monitored Rollback for Coherence in NMP}
The intent is to implement a coherence model for NMP architecture that doesn't hamper the advantages gained and allows simultaneous execution of the  PE in NMP and host CPU. This section briefly discusses the proposed model, whose key features are (1) Minimal off-chip traffic due to the coarse-grained approach and (2) Minimal re-execution on the NMP side. 

\begin{figure}
    \centering
    \includegraphics[width=\textwidth]{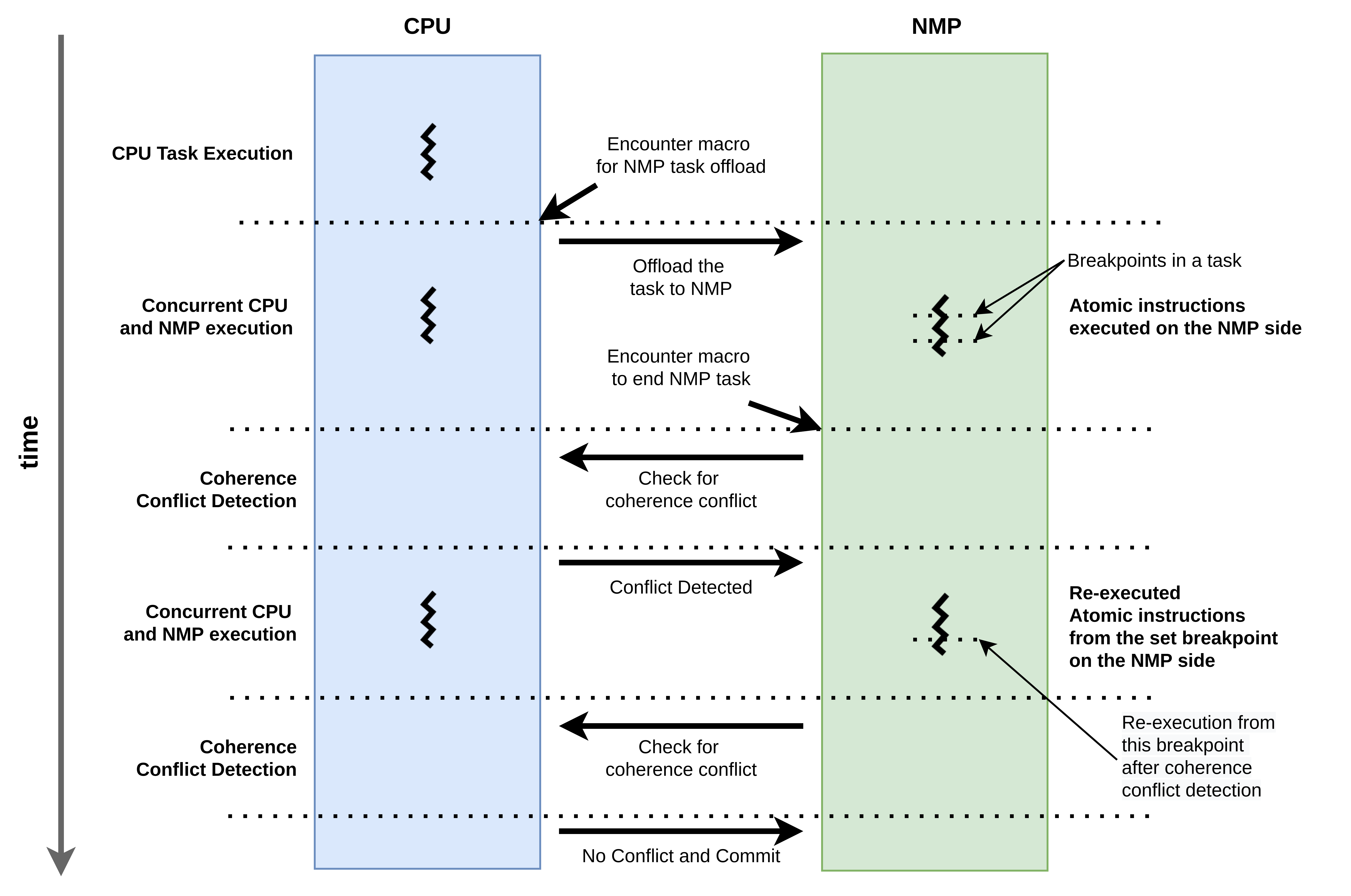}
    \caption{Timeline diagram for Monitored Rollback points based Coherence Mechanism}
    \label{fig:mrcn_timeline}
\end{figure}

The burden is still on the programmer side to define the NMP offloading regions. This we have done in our work using hard macros and the parallelism in tasks is achieved using the OpenMP programming model [7]. Figure \ref{fig:mrcn_timeline} presents the proposed strategy using the timeline diagram. Here, we define the rollback points marked within the code block which is offloaded for execution on to the NMP side.

In equation \ref{LazyPIM_exe_time}, we identify that even under the optimistic execution paradigm as suggested in [2,3], whenever the NMP side experiences a conflict the entire block has to re-executed. We realize the possible scope of improvement with minimal overheads. In MRCN, instead of rolling back the entire set, rollback happens from the first conflict point. This is done by defining breakpoints in the offloaded task. These breakpoints may or may not be equally spaced. We use a priority encoder to select the first conflict point to rollback even if multiple conflicts are detected in the task. So the instructions that were executed before the conflict are not re-executed unnecessarily. The CPU side executes as usual, with the CPU cores coherent in the traditional fine-grained approach. The only additional task is to record the writes done before the validation. Once the validation is done, it erases the previous history and fills it with the new ones. The conflict identification can be marked with tag bits. The coherence message sent from the NMP side is a compressed signature that uses a space-efficient data structure such as the bloom filter used in [2] to contain a large number of elements and a fixed number of bits.

\begin{figure} [!htp]
    \centering
    \includegraphics[width=0.75\textwidth,keepaspectratio]{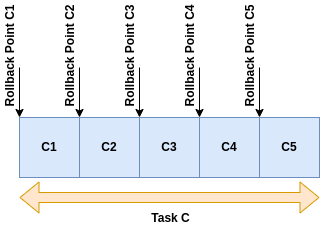}    \caption{Rollback points in MRCN}
    \label{roll}
\end{figure}

  
  In fig. \ref{roll} certain task C is represented as a block which is subdivided into 5 sections from C1 to C5. If the first conflict happens in any of these sections, then the task will rollback to the point meant for that section.

\subsection{Formulating the MRCN strategy}

All the definitions from section \ref{formulating LazyPIM} are consistent in this section as well. As mentioned in the MRCN strategy we consider breakpoints within any offloaded block. Let us assume that there are $b$ breakpoints within any offloaded block $B_i$.

The case for no conflict will 
be same as shown in equation \ref{no conflict LazyPIM}. However, our point of interest is the case when there are coherence conflicts. Let $P^i_{j}$ denote the probability of conflict in between any two breakpoints of $B_i$. The understanding of $P^i_{j}$ is similar to $P^i$ defined in section \ref{formulating LazyPIM}. The only difference between $P^i_{j}$ and $P^i$ is instead of whole block, the probability of conflict is identified between breakpoints $j$ and $j+1$ of block $B_i$. 
\begin{align}
        P^i_{j}=1-[1-\{1-(1-\frac{1}{\mathcal{K}})^{\frac{|\mathcal{N}_i|}{b}}\}\{1-(1-\frac{1}{\mathcal{K}})^{|\mathcal{C}_i|}\}]^\mathcal{K}
\end{align}

Note that upon re-execution from any given breakpoint, the number of instructions for re-execution changes. Therefore, the probability of conflict for subsequent re-execution changes. Let, $P^i_{j,k}$ denote the probability of collision on subsequent re-executions, where $k=j,j+1,\dots b-1$. To further elaborate $P^i_{j,k}$ denotes the probability of conflict when the conflict is received and the rollback happens from the rollback point indexed with $j$ of the offloaded block $B_i$ and after re-execution the conflict is again recieved and this time the rollback happens from the breakpoint k.
We model the average execution time of $B_i$, when conflicts occur. Let us assume that the conflicts are observed between the second and the third breakpoints. Then, the time spent only in the re-execution of instructions from the second breakpoint can be expressed as $\frac{(b-2)\theta_i T_{inst}}{b}$. Then a coherence transaction taking $T_{tran}$ time occurs. Let $\beta_k$ denote $\frac{(b-k)\theta_i T_i}{b}+T_{tran}$. Note that, as the re-execution happens from different breakpoints on the NMP side, the memory addresses accessed by the CPU denoted as a set $\mathcal{C}_i$ also changes. We recognize this variation and denote it using $\mathcal{C}_k$, where $\mathcal{C}_k$ is represented as  $\frac{\beta_j f_{cpu}}{T_{cpu}}$. Therefore, $P^i_{j,k}$ can be expressed as,    

\begin{align}
    P^i_{j,k} = 1-[1-\{1-(1-\frac{1}{\mathcal{K}})^{\frac{|\mathcal{N}_i|}{b}}\}\{1-(1-\frac{1}{\mathcal{K}})^{|\mathcal{C}_k|}\}]^\mathcal{K} 
\end{align}





 Now, we realize the average execution time of $B_i$ when a single coherence conflict is detected,

\begin{align}
    T^{MRCN}_{\substack{\text{with  coherence}\\\text{conflict in } B_i}} =\alpha&+T_{commit}+P^i_{1}\beta_0 \\ \nonumber
    &+P^i_{2}\beta_1 \dots +P^i_{b}(\beta_{b-1})
\end{align}

Now in the case multiple coherence conflicts upon re-execution of $B_i$.
Upon further solving a closed form equation for the average execution time of the MRCN strategy can be reached which is presented in equation \ref{mrcn_final_exe_time}

\begin{align}
\label{mrcn_final_exe_time}
    \mathbb{E}[T^{MRCN}_{i}]&=\alpha+T_{commit} + \sum_{j=1}^b P_j^i \beta_{j-1}
\end{align}


The mean total execution time for the MRCN strategy can be written as $\mathbb{E}[!htp]=\sum_i^N \mathbb{E}[T_i^{MRCN}]$

   




\section{Methodology} \label{methodology}
\begin{table}[!htp]
    \centering
    \caption{CPU Configuration}
    \begin{tabular}{|c|c|} \hline
         CPU & 4 cores, Out-of-Order, 3GHz\\ \hline
         L1I & Private, 32KB, 4-way assoc \\ \hline
         L1D & Private, 32KB, 8-way assoc \\ \hline
         L2 & 4 core Shared, 256KB, 8-way assoc \\ \hline
         L3 & Shared, 16MB, 16-way assoc \\ \hline
         TLB & I-TLB and D-TLB: 256 entries each \\\hline Coherence & MESI Protocol \\
         \hline
         
    \end{tabular}
    
    \label{CPUSpec}
\end{table}
\begin{table}[!htp]
    \centering
    \caption{NMP Configuration}
    \begin{tabular}{|c|c|} \hline
         NMP-CPU & 1 core, In-Order, 2GHz\\ \hline
         L1I & 16KB, 4-way assoc \\ \hline
         L1D & 16KB, 4-way assoc \\ \hline
         L2 & 128KB, 8-way assoc \\ \hline
         L3 & 4MB, 16-way assoc \\ \hline
        TLB & I-TLB and D-TLB: 128 entries each \\ \hline
        Topology & 16 node Mesh \\\hline 
        Coherence & MESI Protocol \\
        \hline
    \end{tabular}
    
    \label{NMPSpec}
\end{table}
\begin{table}[!htp]
\centering
 \caption{3D Memory Configuration}
\begin{tabular}{|l|l|}
\hline
Bankgroups              & 4   \\ \hline
Banks per group         & 4   \\ \hline
Device Width (bits)     & 128 \\ \hline
Capacity (GB)           & 4   \\ \hline
Number of Channels      & 8   \\ \hline
Channel Size (MB)       & 512 \\ \hline
\end{tabular}%
\label{HBMSpec}
\end{table}
\begin{figure*} [!htp]
    \centering
    \subfigure[Granularity:100]{\label{lazy_vs_analytic}\includegraphics [width=0.475\textwidth]{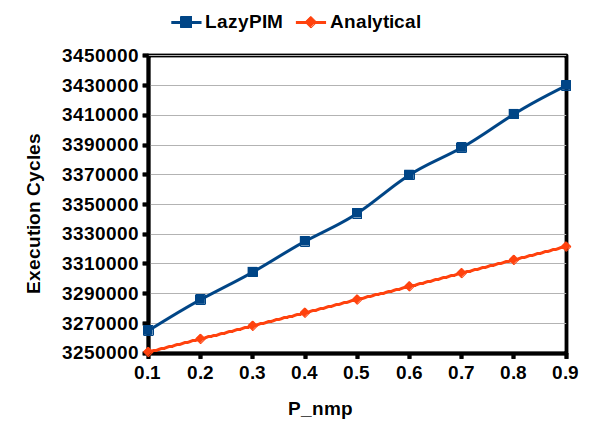}}
    \subfigure[Granularity:500]{\label{lazy_vs_analytic500}\includegraphics [width=0.475\textwidth]{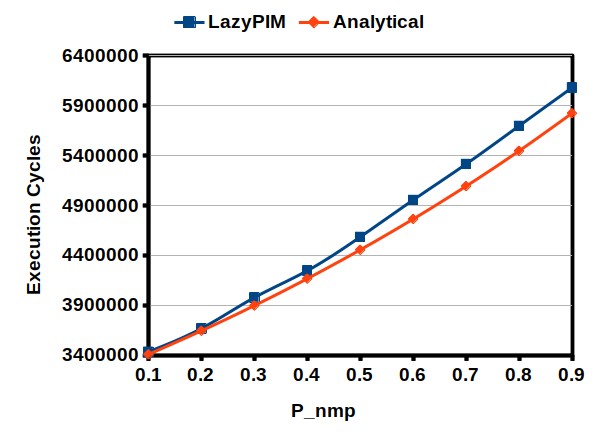}}
    \subfigure[Granularity:100]{\label{error_in_analytic}\includegraphics [width=0.475\textwidth]{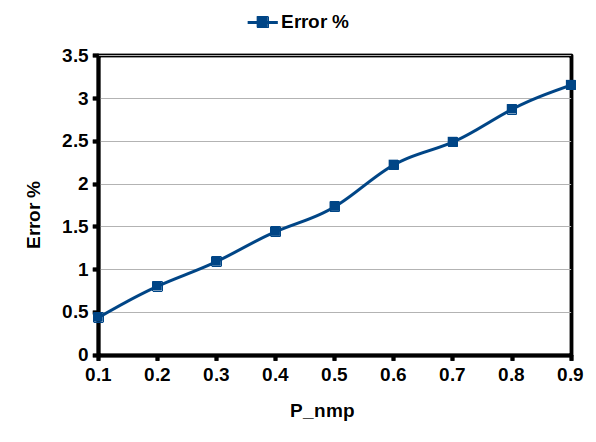}}
    \subfigure[Granularity:500]{\label{error_in_analytic500}\includegraphics [width=0.475\textwidth]{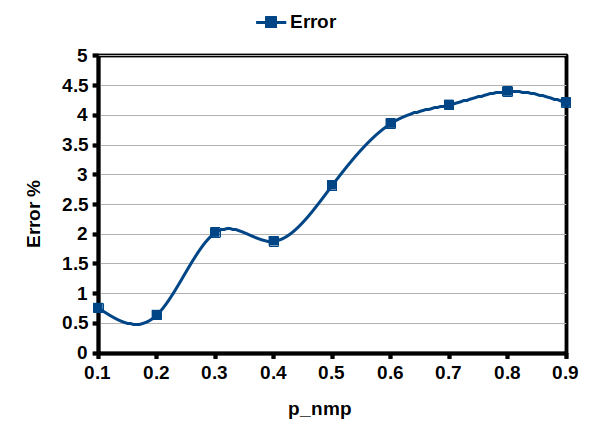}}
    \caption{Comparison of the total execution time between the analytical and simulated results for CONDA coherence strategy with variations in the percentage of shared memory accessed by the NMP, $f_{nmp}$. Each plot is for different offloaded block granularities. Subsequent Error\% between the analytical and simulated results for CONDA are also presented.}
    \label{lazy_analytical_error} 
\end{figure*}
Our simulation consists of a multicore CPU with private L1(I) \& L1(D) caches, partially shared L2 caches and a shared L3 cache. The CPU specification details are presented in table \ref{CPUSpec}. We have profiled the performance for HBM as the main memory unit [18], the configuration details are presented in table \ref{HBMSpec}. NMP cores used are simple in-order core, with all caches are private to the individual core. The details for the NMP specifications are presented in table \ref{NMPSpec}. 

We have emulated the existing CPU-NMP coherence strategies presented in [2,3], after completely profiling the system behaviour. For profiling the system performance, we have used Structural Simulation Toolkit (SST) 11.1.0 [24], DRAMSim3 [17] and Valgrind [20]. To further validate the performance of our emulated enviroment, we execute the same set of benchmarks on MultiPIM [27] for the same NMP architecture as shown in fig. \ref{nmp_baseline_diagram}. We generate synthetic traces to perform the study of our model in detail. The advantage of working upon synthetic benchmarks at first is the added controlability to a number of factors. The system therefore can be tested from the best possible scenario to the worst case scenario. To further substantiate the work, we then perform the evaluation against the real world benchmarks from the Rodinia Benchmarks Suite (RBS) [6]. Following four real-world benchmarks from RBS have been examined - . The execution cycles is the observed metric in the results evaluation, since, coherence message transactions effect the system performance in terms of execution cycles. 

\section{Results} \label{results}
In this section, we study the analytical and simulated environment of a prominent existing coherence strategy- CONDA and examine the improvement achieved by MRCN. The section is divided into two parts, with first part focusing on the synthetic benchmarks and the second part studying the real-world benchmarks from RBS [6].

\begin{figure*} [!htp]
    \centering
    \subfigure[Granularity:100]{\label{MRCN_vs_analytic100}\includegraphics [width=0.475\textwidth]{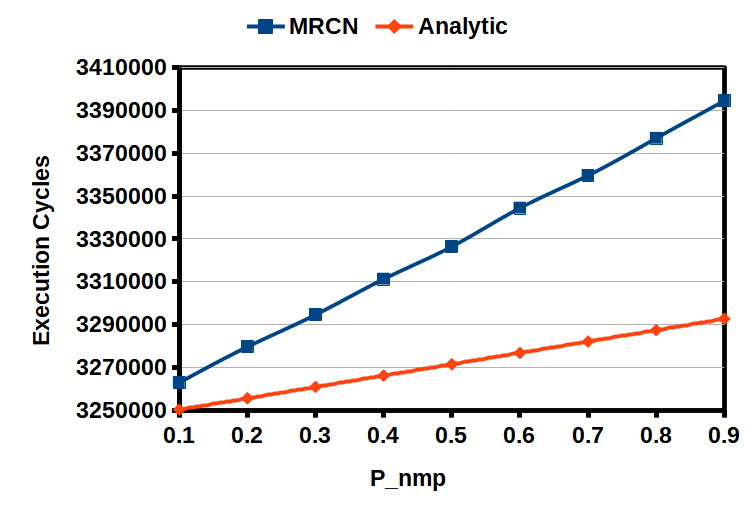}}
    \subfigure[Granularity:500]{\label{MRCN_vs_analytic500}\includegraphics [width=0.475\textwidth]{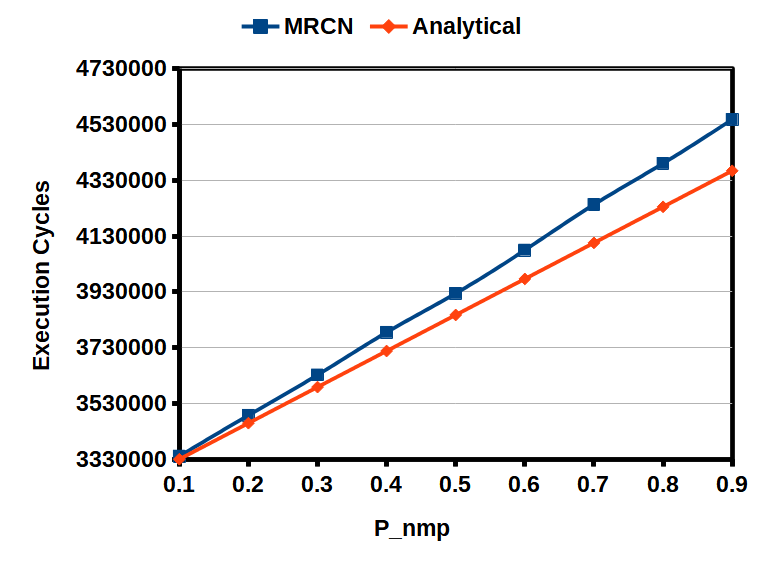}}
    \subfigure[Granularity:100]{\label{error_in_MRCNanalytic100}\includegraphics [width=0.475\textwidth]{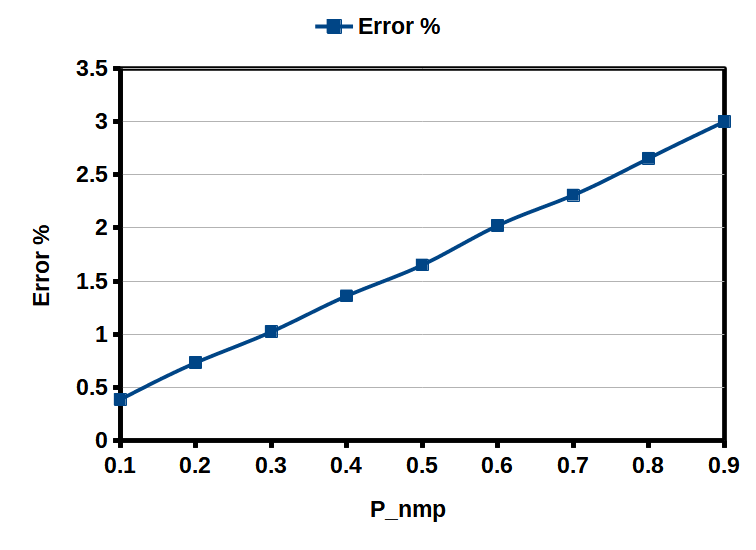}}
    \subfigure[Granularity:500]{\label{error_in_MRCNanalytic500}\includegraphics [width=0.475\textwidth]{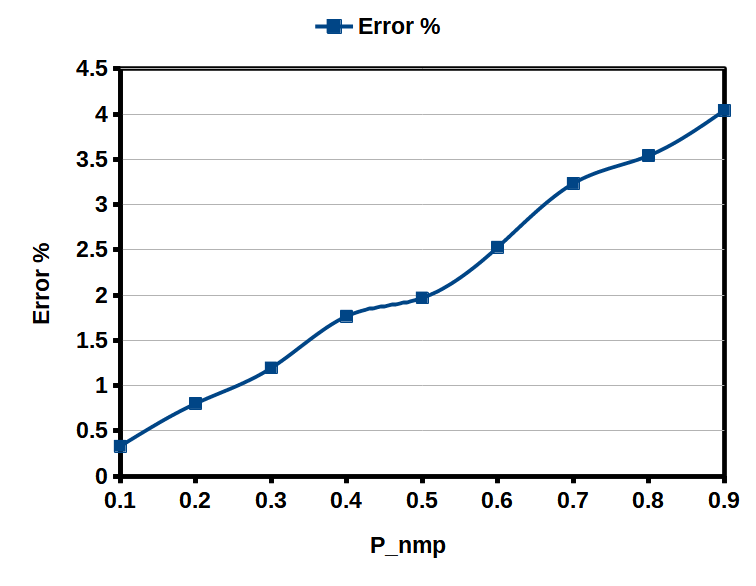}}
    \caption{Comparison of the total execution time between the analytical and simulated results for MRCN coherence strategy with variations in the percentage of shared memory accessed by the NMP, $f_{nmp}$. Each plot is for different offloaded block granularities. Subsequent Error\% between the analytical and simulated results for MRCN are also presented.}
\end{figure*}
\subsection{Study with Synthetic Benchmarks}
We validate our analytical results with the simulated results and present the performance improvement achieved by the MRCN strategy. 

\subsection{Evaluating CONDA Analytical Model}

In fig. \ref{lazy_analytical_error}, we compare the analytical results for the benchmark execution time of the CONDA strategy as obtained from equation \ref{LazyPIM_exe_time}. Our examination in fig. \ref{lazy_vs_analytic} and fig. \ref{lazy_vs_analytic500} reveals that for a granularity of 100 and 500 instructions executed in the offloaded code block on the NMP side, respectively, equation \ref{LazyPIM_exe_time} succeeds in giving the approximate execution time with an accuracy of $3.2\%$ or less. As defined in section \ref{analyticalmodel}, $f_{nmp}$ depicts the percentage of the shared memory accessed by the NMP to complete the benchmark. $f_{cpu}$ is kept fixed at $0.5$, as our objective is to increase the reliability on the NMP side. As we increase, $f_{nmp}$, the probability of conflict $P^i$ also increase as indicated in equation \ref{LazyPIM_conflict_prob}. The increase in $P^i$ contributes to multiple re-execution of the complete offloaded code block. This inherently increases the total execution time. It is worth mentioning that a lot of effort is given to having lower coherence conflicts. Hence, from our examination, we realize that the cases where $f_{nmp}$ is lower than 0.5 are more likely to happen than the cases where $f_{nmp}$ is greater than 0.5. However, having the advantage of operating over synthetic benchmarks, we perform an adversarial evaluation. As shown in fig. \ref{error_in_analytic} and fig. \ref{error_in_analytic500} for a granularity of 100 and 500 instructions respectively, for lower values of $f_{nmp}$ highly accurate execution time values are available from the equation \ref{LazyPIM_exe_time}, with error within $2\%$ or less. 

\subsection{Evaluating MRCN model}

In fig. \ref{MRCN_vs_analytic100} and fig. \ref{MRCN_vs_analytic500}, we plot the execution time from the simulations and the expected execution time from the analytical model for granularity of 100 and 500 instructions, respectively, for the individual offloaded code blocks against varying the percentage of shared memory space used by the NMP PEs, $f_{nmp}$. Similar to the case of CONDA modeling and evaluation, the percentage of shared memory space used by CPU side $f_{cpu}$ is fixed. Fig. \ref{error_in_MRCNanalytic100} and fig. \ref{error_in_MRCNanalytic500} show the error percentage between the analytical and simulated execution times. The expected execution time for the MRCN strategy, as shown in equation \ref{mrcn_final_exe_time}, with $4\%$ error or less, predicts the execution time when the benchmark is put on simulation.
Reasoning similar to the case of CONDA stated in the previous paragraph holds for the MRCN strategy as well. Despite the added feature of breakpoints, increasing the $f_{nmp}$ causes an increase in the number of coherence conflicts and hence the re-execution of offloaded code blocks to ensure the reliability of execution. However, the breakpoints reduce the unnecessary time spent re-executing the entire offloaded code block and promote partial code block re-executions from the most relevant breakpoint. 

\subsection{Evaluating CONDA against MRCN}
\begin{figure} [!htp]
    \centering
    \subfigure[Granularity:100]{\label{exe_a_pnmp}\includegraphics [width=0.495\textwidth]{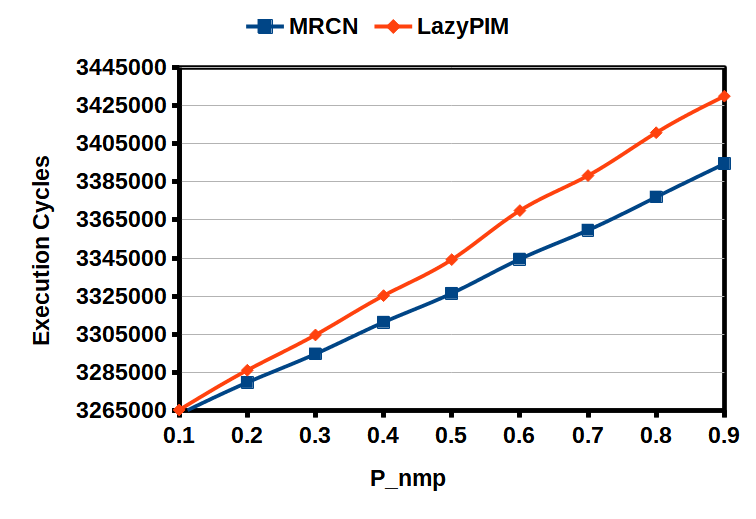}}
    \subfigure[Granularity:500]{\label{exe_a_pnmp500}\includegraphics [width=0.495\textwidth]{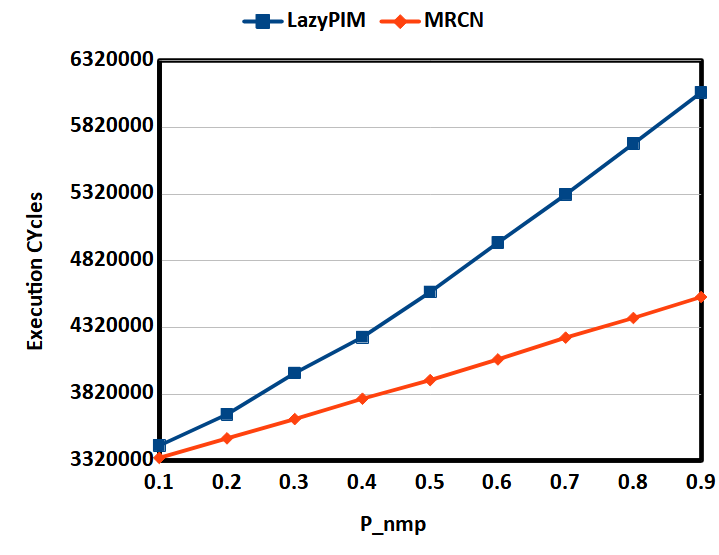}}
    \caption{Improvement in Execution time between CONDA and MRCN, observed across two different block granularities}
\end{figure}

One of the major contibutions of this study is to propose a more enhanced coherence mechanism. We have proposed and compare the performance of the MRCN strategy with a recognized existing work- CONDA [2,3]. Fig. \ref{exe_a_pnmp} and fig. \ref{exe_a_pnmp500} present a direct comparison of the two coherence mechanisms in terms of execution times across two different code block granularities. For synthetic benchmarks where the $f_{nmp}$ is very low, MRCN mechanism performs little better than CONDA.  As $f_{nmp}$ increases the execution time for the MRCN strategy is relatively very less compared to the CONDA strategy. One of the key reasons behind the major performance improvement is the identification of code sections within each offloaded code blocks and avoiding the re-execution of complete code block.

\begin{figure} [!htp]
    \centering
    \subfigure[$f_{nmp}$: 0.1]{\label{exe_pnmp}\includegraphics [width=0.495\textwidth]{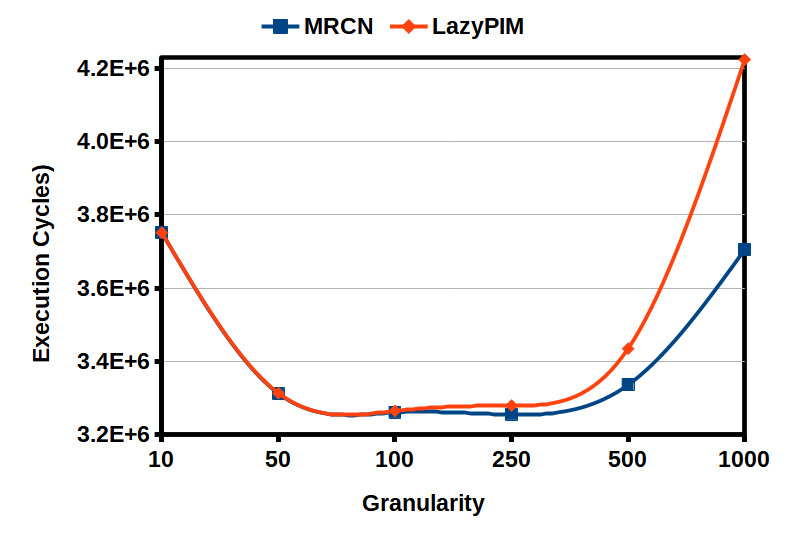}}
    \subfigure[$f_{nmp}$: 0.9]{\label{exe_pnmp500}\includegraphics [width=0.495\textwidth]{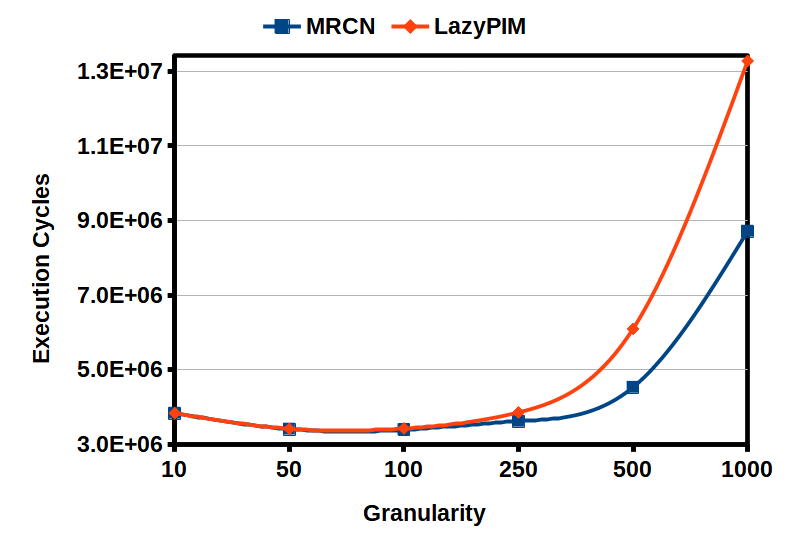}}
    \caption{Comparison in performance of CONDA and MRCN across different granularities for two different $f_{nmp}$}
    \label{synth_gran}
\end{figure}

In fig. \ref{synth_gran}, we present a comparative plot between the CONDA and MRCN coherence strategies by varying the granularities. In fig. \ref{exe_pnmp}, we fix the percentage of shared memory used by the NMP side to 0.1, and in fig. \ref{exe_pnmp500}, we set the percentage shared memory used by NMP as 0.9 and sweep the granularities. In both cases, it is visible that the MRCN, because of the inclusion of breakpoints and avoidance of unnecessary re-execution, performs far better than the CONDA strategy. Another significant observation from fig. \ref{exe_pnmp} and \ref{exe_pnmp500}, is that for certain granularities, such as for 100 and 250 instruction granularities, the system performs exceedingly well irrespective of the impact of the strategy. These granularities are in between the two extremes. One extreme is highly fine-grained, and the other is offloading huge blocks for execution, hence providing suitable opportunities for coherence checks and, if needed, block re-executions.


\begin{figure} [!htp]
    \centering    \subfigure[Granularity:100]{\label{speedup100}\includegraphics [width=0.8\textwidth]{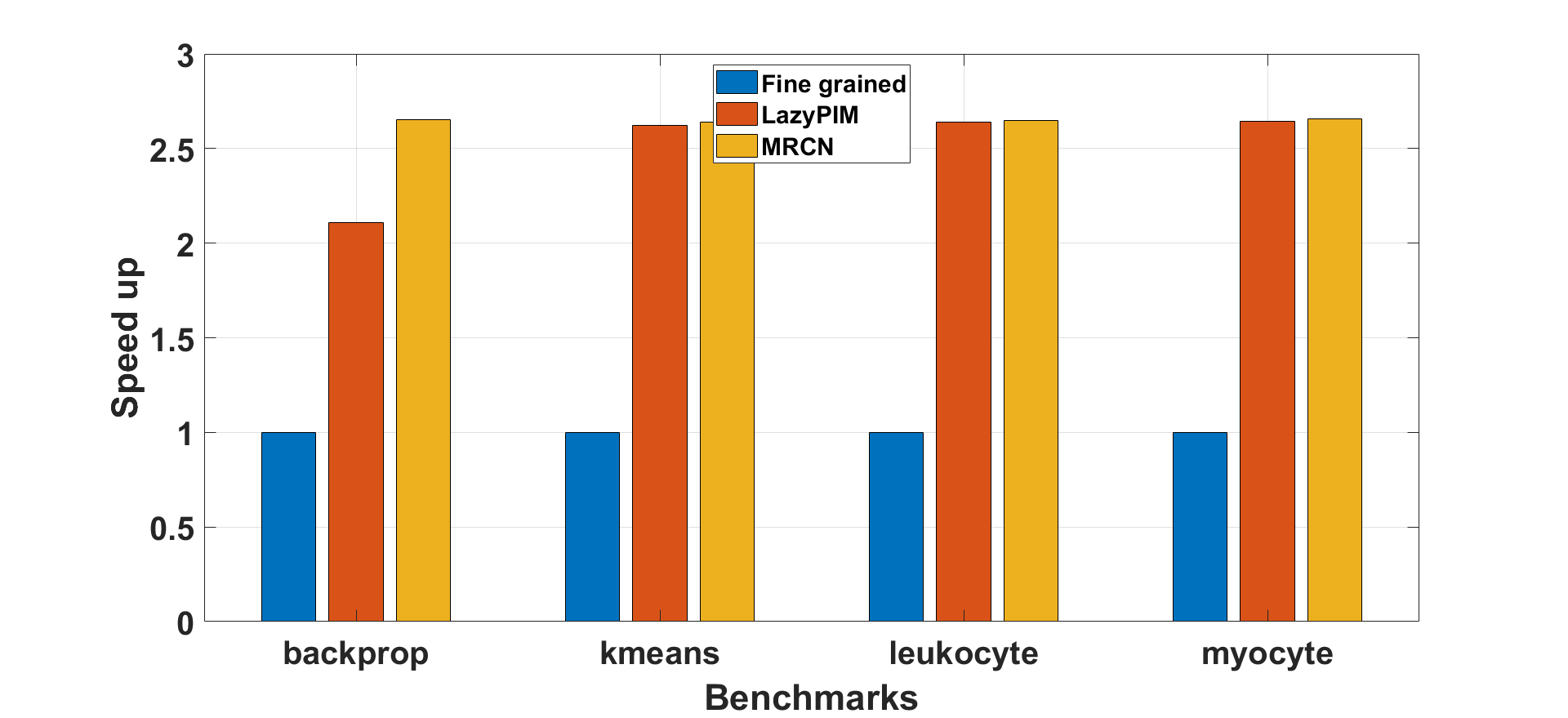}}
    \subfigure[Granularity:500]{\label{speedup500}\includegraphics [width=0.8\textwidth]{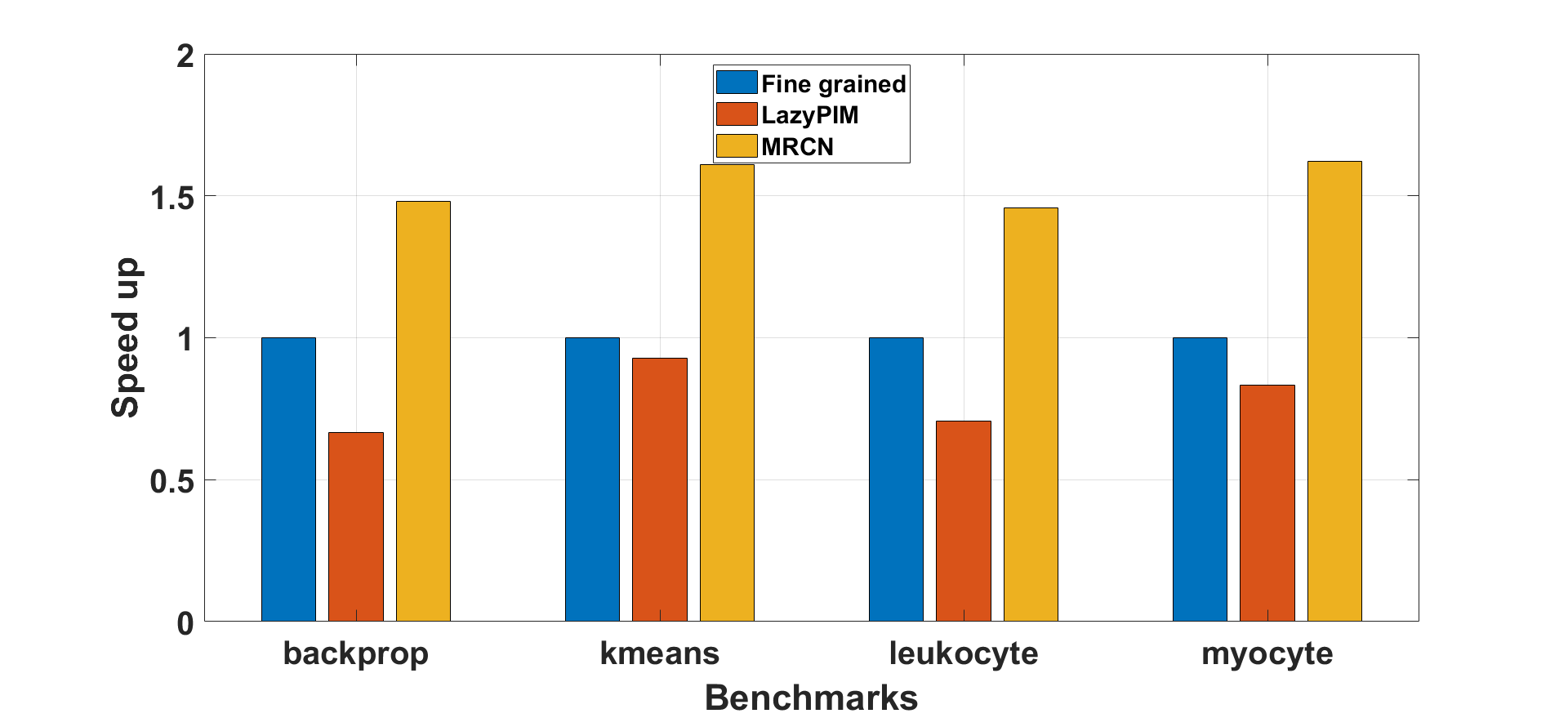}}
    
    \caption{Normalised Speed-up for different coherence strategies against real world benchmarks from RBS [6]}
    \label{real_world_becnhmarks_speedups}
\end{figure}
\subsection{Study with Real-World Benchmarks}

In fig. \ref{real_world_becnhmarks_speedups}, we study the speedup achieved for four of the benchmarks from the RBS [6], evaluated against three different benchmarks. As discussed in section \ref{nmp_task_offload}, the code blocks which thrive on data locality are executed on the CPU, and maximum benefits of parallelization are obtained from the non-cacheable code blocks offloaded to the NMP side. In fig. \ref{real_world_becnhmarks_speedups}, we perform the study for two granularities of offloaded code blocks which are 100 and 500 instructions, shown in fig. \ref{speedup100} and fig. \ref{speedup500} respectively. One of the major take away from the fig. \ref{real_world_becnhmarks_speedups} is that the CONDA performs better for lower granularity compared to fine-grained coherence, as it achieves a speedup of almost two times compared to that of the fine-grained. Whereas the MRCN, irrespective of the granularity of the offloaded code block, shows consistent improvement in speedup across both the granularities. One of the key reasons behind the performance degradation of CONDA, when subjected to larger granularities, is the re-execution of the complete tasks in the code block; this adds significantly to the execution time. For MRCN, since the breakpoints hold the information of the last point from which the rollback has to be initiated. With a nominal hardware overhead to hold the information about the rollback points, the NMP PE avoids executing the complete code block and re-executes only the necessary portions. Hence, MRCN can achieve much better performance than CONDA and Fine-grained coherence mechanisms.

\section{Conclusion \& Future Work} \label{conclusion}

Maintaining coherence between CPU and NMP impacts the overall system performance as each unit operates on shared memory space. Most of the literature works by adopting one of the strategies, i.e., either restricting the memory space for any single processing unit to be used or using the same coherence model as used in a multicore architecture. However, we realize that addressing the coherence issue in such systems and proposing suitable improvements is a significant research problem, as both the system's performance and reliability are affected. We have successfully modeled the CONDA strategy analytically and identified the potential improvements. We then proposed MRCN, which consistently showed better performance than the CONDA strategy, with a minimum of 0.19ms time saved in execution, which is significant as the systems operate at GHz frequency. We hope that our analytical model gives a clear picture of all the operations performed in maintaining coherence and motivates more work to come on the same.  

\includepdf[pages=-]{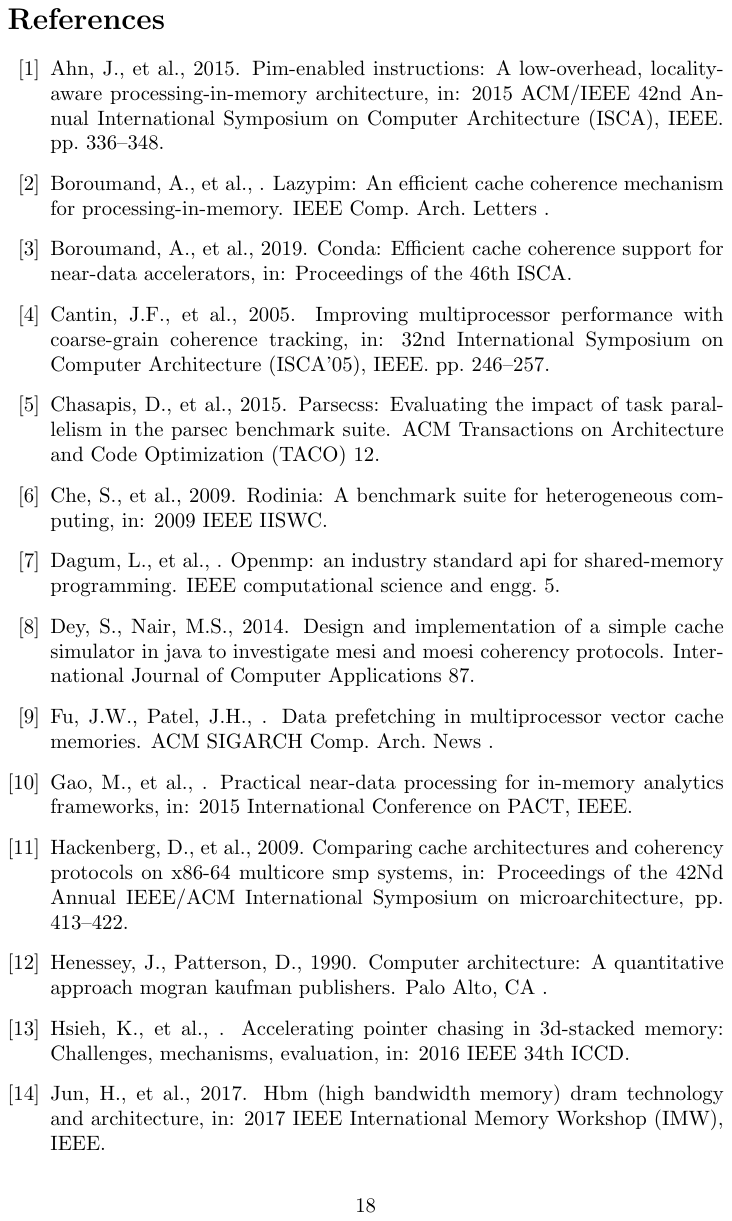}

\end{document}